\documentclass[twocolumn,showpacs,preprintnumbers,amsmath,amssymb]{revtex4}

\usepackage{graphicx,subfigure,epsfig,epstopdf}
\usepackage{amsmath,amssymb}
\input{epsf}
\input{epsfx}

\newcommand{\oad}{\mathbf{a}^{\dagger}}

\newcommand{\oa}{\mathbf{a}}

\def\ket#1{\mathinner{|{#1}\rangle}}
\def\braket#1{\mathinner{\langle{#1}\rangle}}

 \def\braket#1{\mathinner{\langle{#1}\rangle}}

\begin{document}
\title{Coherent generation of nonclassical light on a chip via photon-induced tunneling and blockade}

\author{Andrei Faraon$^{1 *}$, Ilya Fushman$^{1 *}$, Dirk Englund$^{1 *}$, Nick Stoltz$^{2}$, Pierre Petroff$^{2}$ \& Jelena Vu\v{c}kovi\'{c}$^{1}$ \footnote[2]{Correspondence and requests for materials should be addressed to J. Vuckovic (email: jela@stanford.edu)}\linebreak
{\it $^{*}$These authors contributed equally to this work.}\linebreak}

\affiliation{$^{1}$E. L. Ginzton Laboratory, Stanford University, Stanford CA 94305\\
}

\affiliation{$^{2}$Department of Electrical and Computer Engineering, University of California, Santa Barbara, CA 93106}

\begin{abstract}
We report the observation of nonclassical light generated via {\it photon blockade} in a photonic crystal cavity with a strongly coupled quantum dot. By tuning the frequency of the probe laser with respect to the cavity and quantum dot resonance we can probe the system in either {\it photon blockade} or {\it photon-induced tunneling} regime. The transition from one regime to the other is confirmed by the measurement of the second order correlation that changes from anti-bunching to bunching.

\end{abstract}

\maketitle

Quantum dots in photonic crystals are interesting for their potential in quantum information processing\cite{98LossQDquantcomp, 99ImamogluQDquantcomp} and as a testbed for cavity quantum electrodynamics. Recent advances in controlling\cite{06ImamQDspinPrep} and coherent probing\cite{NatureRef,2007KartikPainterNature} of such systems open the possibility of realizing quantum networks originally proposed for atomic systems\cite{CZKM1997PRL, 2006.Kimble.Quant.Net, 2007MonroeAtomEntanglement}. We recently showed\cite{NatureRef} that the intensity of a probe beam coupled to a photonic crystal cavity is controlled by the presence of a single strongly coupled quantum dot that splits the cavity resonance into two eigenstates\cite{SCImamogluNature, Yoshie04}. Here we analyze the photon statistics of this probe beam. We show that the capture of a single photon into the cavity affects the probability that a second photon is admitted. This probability drops when the probe is positioned at one of the two energy eigenstates, resulting in photon antibunching. This result is analogous to the recent report on photon blockade of a neutral atom strongly coupled to a Fabry-Perot cavity\cite{KimbleBlockade,97ImamogluBlockade}. In addition, we show that when the probe is positioned in between the energy eigenstates the probability of admitting subsequent photons increases, resulting in photon bunching. We call this process {\it photon-induced tunneling}. This system represents an ultimate limit for solid-state nonlinear optics at the single photon level. Along with demonstrating the generation of nonclassical photon states, we propose an implementation of a single photon transistor\cite{LukinSPhTr} in this system.

\begin{figure}[htp]
    \includegraphics[width=3.5in]{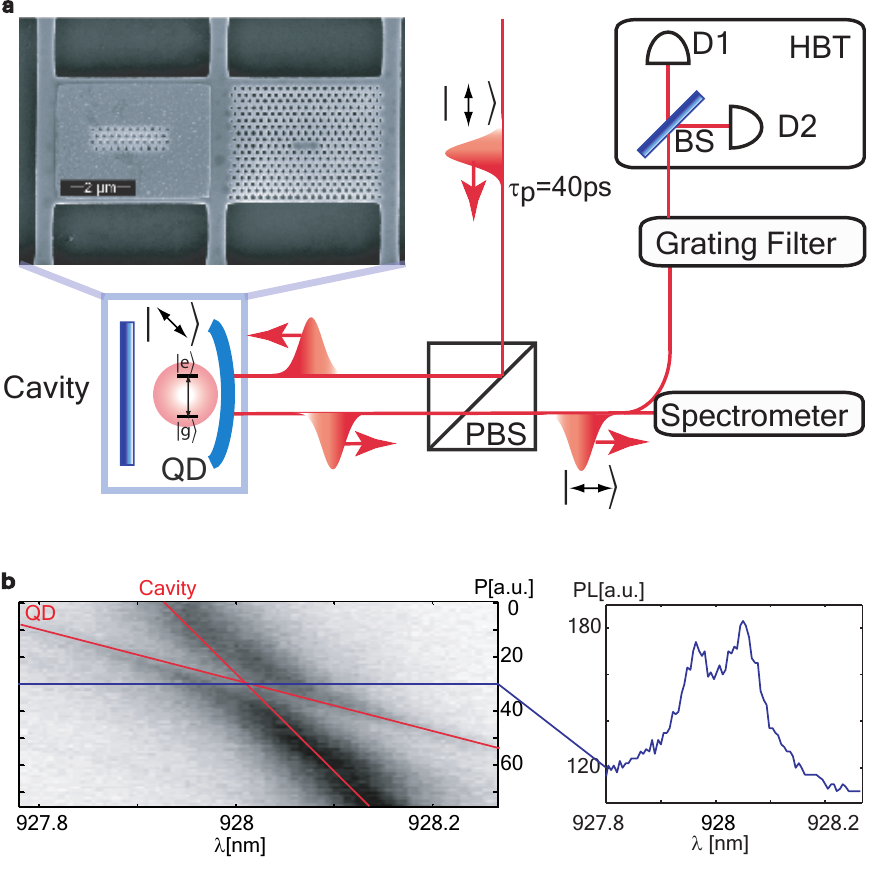}
    \caption{\footnotesize(a) Experimental setup. Laser pulses (40ps FWHM) are reflected from a photonic crystal cavity linearly polarized at $45^{\circ}$ to the input polarization set by the polarizing beam splitter (PBS). The output light, observed in cross-polarization and carrying the cavity-coupled signal, was analyzed using a Hanbury-Brown-Twiss (HBT) setup that measures second-order correlation. The inset shows the suspended structure with the photonic crystal cavity and the metal pad for local temperature tuning\cite{AndreiTtune}. (b) Anticrossing observed in photoluminescence as the QD is tuned into resonance with the cavity. The temperature tuning is done by linearly increasing the power P of the heating laser\cite{AndreiTtune}. The red lines mark the cavity and QD resonance if they were decoupled. The right panel shows the spectrum at the anticrossing point marked by the blue line.} 
    \label{fig:setup}
\end{figure}

The optical system consists of a self-assembled InAs quantum dot (QD) with decay rate $\gamma/2\pi=0.1$ GHz, coupled to a three-hole defect cavity\cite{NodaL3} in a two-dimensional GaAs photonic crystal, as described in ref.\cite{NatureRef}. The coupling rate $g/2\pi = 16$ GHz equals the cavity field decay rate $\kappa/2\pi = 16$ GHz (corresponding to a cavity quality factor Q=10,000), putting the system in the strong coupling regime\cite{Yoshie04, SCImamogluNature}. We first characterize the system in photoluminescence (PL) by pumping the structure above the GaAs bandgap.  The PL scans in Fig.\ref{fig:setup}(b) show the anticrossing characteristic of strong coupling between the QD and the cavity. Here, the QD is tuned into resonance using local temperature tuning\cite{AndreiTtune} around an average temperature of 20 K maintained in a continuous flow cryostat. We probe the system with linearly polarized laser beams (Fig.\ref{fig:setup}(a)) and observe the cross-polarized output as described in our previous work\cite{NatureRef}. The measurement on the reflected port from this single-sided cavity is analogous to a transmission measurement in a Fabry Perot arrangement.

\begin{figure}[htbp]
    \includegraphics[width=3.5in]{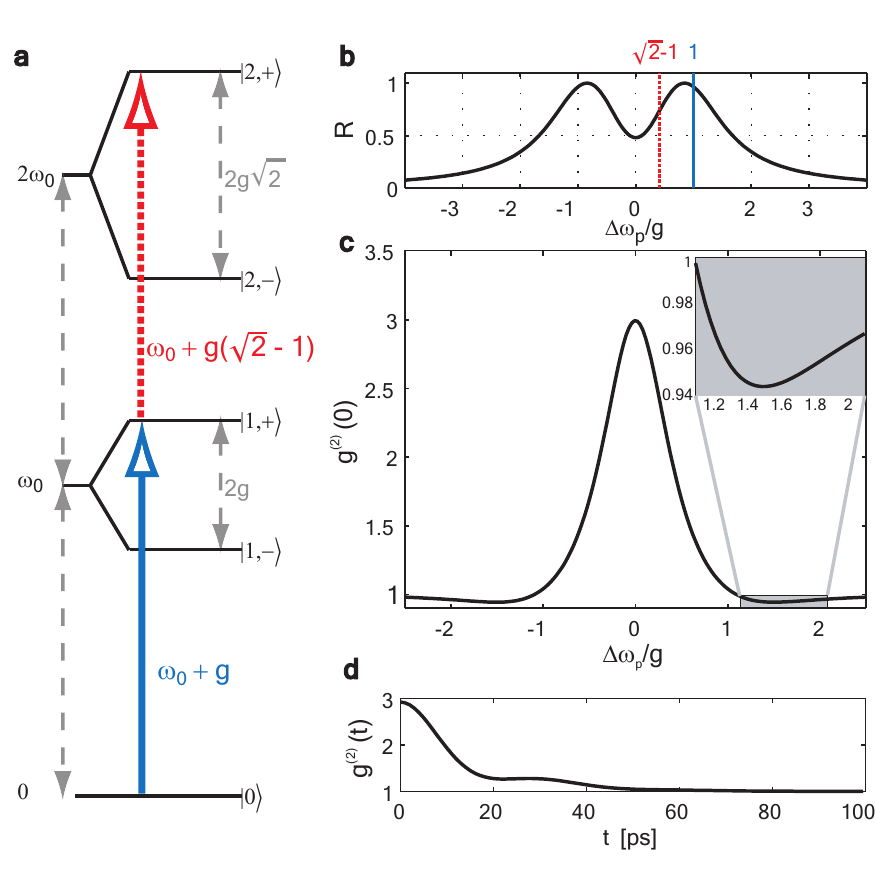}
    \caption{\footnotesize(a) Energy diagram showing the first and second-order manifolds of the strongly coupled cavity/QD system. The energy difference between consecutive manifolds is not constant as shown by the blue and the red arrows. This anharmonic spacing of the levels causes phenomena such as photon blockade\cite{KimbleBlockade} and photon induced tunneling. (b) Simulated output intensity for a probe frequency tuned through the cavity/QD strongly coupled system. The red and blue lines indicate the transitions $\ket{0} \rightarrow \ket{1,+}$ and $\ket{1,+} \rightarrow \ket{2,+}$. Photon blockade is expected at $\Delta \omega _{p}/g \sim 1$ because the absorption of a photon into $\ket{1,+}$ suppresses the probability of absorbing a second photon of the same energy for a transition to $\ket{2,+}$. As $\Delta \omega _{p} \rightarrow 0$ the absorption of a photon into the first manifold enhances the absorption probability into higher order manifolds. (c) Computed second-order correlation $g^{(2)}(0)$ for a coherent laser probe reflected from the cavity. Inset shows that photon blockade is expected when the probe detuning is $\Delta \omega_{p}/g \sim 1.5$ (and not $\Delta \omega_{p}/g \sim 1$) because of the finite linewidth of the polaritons. As $\Delta \omega_{p} \rightarrow 0$ the output field is bunched. (d) Simulated time dependence of the second-order correlation for $\Delta \omega _{p} = 0$. The value for ($g^{(2)}(\tau)$) drops rapidly for time delays greater than $\sim 5$ps, corresponding to the cavity photon lifetime.} 
    \label{fig:theory}
\end{figure}

The energy eigenstates of a two-level system strongly coupled to an optical resonator are grouped into two-level manifolds denoted $\ket{\pm,n}$, with energies $\hbar \omega _{n,\pm}=\hbar(n \omega_{0} \pm g \sqrt{n})$ , where $n$ is the number of energy quanta in the system and $\omega_{0}$ is the bare cavity frequency (Fig.\ref{fig:theory}(a)). The anharmonic energy level spacing causes phenomena such as photon blockade\cite{KimbleBlockade} or photon-induced tunneling. To observe photon blockade, a coherent probe beam (frequency $\omega _{p}$) tuned at $\omega _{1,\pm}=\omega_{0} \pm g$ is coupled to the cavity. This probe is resonant with the first-order manifold, but detuned from transitions to the second manifold, $\omega _{1 \rightarrow 2}=\omega_{0} \pm g(\sqrt{2}\mp 1)$ as shown in Fig.\ref{fig:theory}(a). Consequently, once a photon is coupled into the system, it suppresses the probability of coupling a second photon with the same frequency. As a result, the output field acquires sub-poissonian statistics with second-order correlation $g^{(2)}(0) < 1$, as recently demonstrated by Birnbaumm et al\cite{KimbleBlockade} in an experiment with neutral atoms.

In addition to photon blockade, photon-induced tunneling is expected near the bare cavity resonance ($\omega_{p}-\omega_{0}=\Delta \omega _{p} \rightarrow 0$): the absorption of a first photon enhances the absorption of subsequent photons so the output consists of ``photon bunches.'' In Fig.\ref{fig:theory}(b) we show the output spectrum as the probe is tuned through the cavity and mark the resonance of the transitions $\ket{0} \rightarrow \ket{1,+}$ and $\ket{1,+} \rightarrow \ket{2,+}$. The simulated driving field injects an average photon number $\left\langle n \right\rangle \sim 0.4$ when resonant with the polaritons in the first manifold, thus causing a slight saturation of the QD dipole. As $\Delta \omega _{p} \rightarrow 0$, the probability of absorbing the first photon decreases. However, if a photon is nevertheless absorbed, it enhances the probability of capturing the second photon. This process produces a photon-bunched output. This intuitive explanation can be extended in the same way to include higher order manifolds. 

The expected second-order correlation function for our system is shown in Fig.\ref{fig:theory}(c), where we plot the dependence of $g^{(2)}(0)$ for different detunings $\Delta \omega _{p}$ of the probe  from the anticrossing frequency $\omega _{0}$.  As expected from the intuitive argument above, the simulation predicts photon bunching  as $\Delta \omega _{p} \rightarrow 0$. Photon blockade is evident in the antibunched region near $\Delta \omega _{p} \sim \pm 1.5 g$ (see inset of Fig.\ref{fig:theory}(c)). However, the blockade does not occur for $\Delta \omega _{p}=\pm g$ as previously explained, but at $\Delta \omega _{p} \sim \pm 1.5 g$. The reason is that the linewidth of the eigenstates ($\sim \kappa$) is comparable to the splitting of the manifolds ($\sim 2g$), resulting in significant overlap of the allowed transitions between consecutive manifolds.

\begin{figure*}[htbp]
    \includegraphics[width=5in]{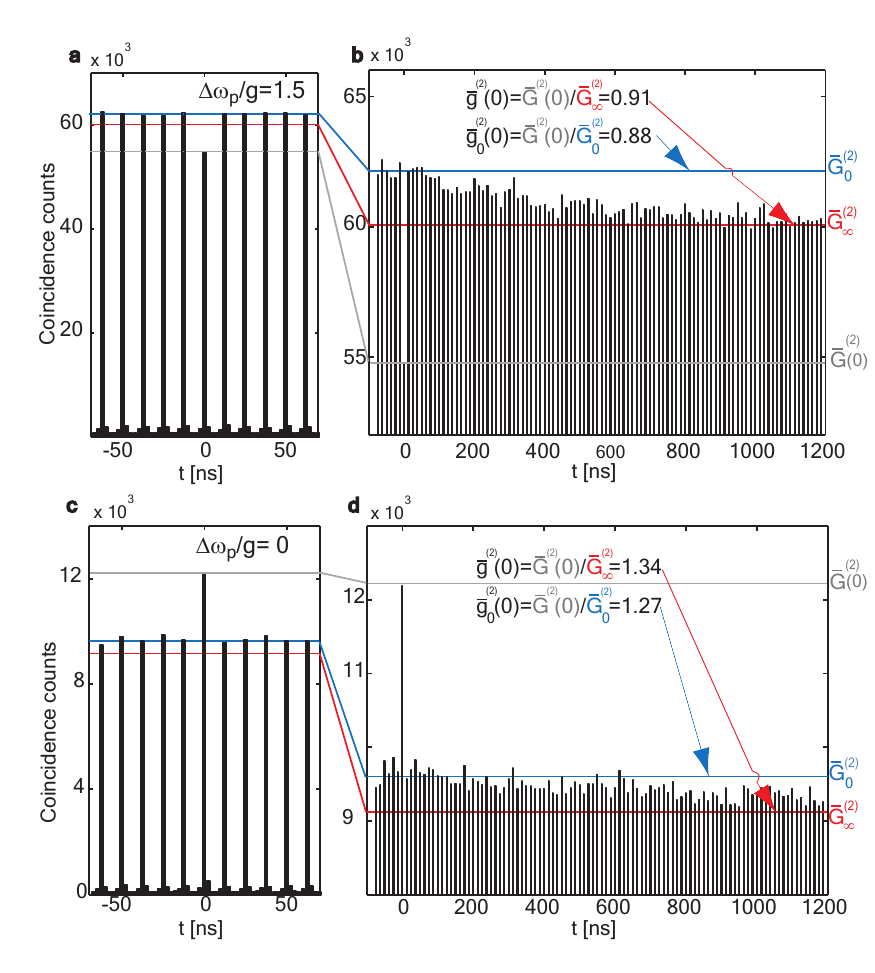}
    \caption{\footnotesize Measurement of the second-order correlation function for coherent laser pulses reflected from the photonic crystal cavity with a strongly coupled QD. (a) Photon blockade, manifested here in the sub-poissonian statistics, observed when the probe is detuned by $\Delta \omega_{p} /g = 1.5$. (b) Beside the antibunching at $\tau = 0 $ the data in panel {\bf a} shows bunching due to QD blinking over timescales of hundreds of nanoseconds. The red and blue lines show the normalization levels $\overline{G}^{(2)}_{\infty}$ and $\overline{G}^{(2)}_{0}$ observed as $\tau \rightarrow \infty$ and as $\tau \rightarrow 0$ respectively. The second-order correlation is $\overline{g}^{(2)}(0,1.5)=0.91$ when normalized to $\overline{G}^{(2)}_{\infty}$ and $\overline{g}_{0}^{(2)}(0,1.5)=0.88$ when normalized to $\overline{G}^{(2)}_{0}$ (c) Photon-induced tunneling is observed when the laser pulse is tuned at the anticrossing wavelength $\Delta \omega_{p} /g = 0$. (d) Depending on the choice of normalization for the data in panel {\bf c}, the second-order correlation takes the values $\overline{g}^{(2)}(0,0)=1.34$ and $\overline{g}_{0}^{(2)}(0,0)=1.27$.}
    \label{fig:antibunch}
\end{figure*}

We measure the time-dependent autocorrelation $g^{(2)}(\tau)$ using the Hanbury-Brown-Twiss (HBT) setup shown in Fig.\ref{fig:setup}(a) and described in ref.\cite{DirkAndreiSPhtransfer, 05PRLEnglund}. The relevant features occur at the QD-cavity coupling rate $g$, enveloped by the coherence time\cite{91Charmichael}, as shown in Fig.\ref{fig:theory}(d). The coherence time for our system is given by the cavity photon lifetime $1/2\kappa$. Hence the time-dependent features in $g^{(2)}(\tau)$ occur much faster than the 300 ps time-resolution of the single photon counting modules in the HBT setup. For this reason, we sample the autocorrelation function by short pulses ($\Delta t_{FWHM} \sim 40$ ps, $\Delta \omega _{FWHM}/2\pi \sim 12$ GHz) with a repetition rate of $12.5$ns. This probe pulse duration forms a compromise between fast sampling and a spectral linewidth narrow enough to resolve the relevant spectral features. In the remainder of the paper we present the measurements of $g^{(2)}(\tau)$ for different detunings of the probe beam, denoted as $g^{(2)}(\tau,\Delta \omega_{p} / g)$.

To observe photon blockade and photon-induced tunneling, we measured the second-order correlation at detunings $\Delta \omega _{p}/g=-1.5$ and $\Delta \omega _{p}/g=0$ as shown Fig.\ref{fig:antibunch}. The expected photon antibunching and bunching behavior is clearly visible at zero time delay (Fig.\ref{fig:antibunch}(b,d)). The histograms also show bunching over timescales of hundreds of nanoseconds.  This bunching results not from the quantum nature of the system, but from blinking behavior of the QD.  As reported by Santori et al\cite{CharlieBlinkingPRB}, such blinking behavior between a {\it bright} and a {\it dark} state results in bunching near t=0, with a characteristic fall-off rate given by the mean switching rate. Our observations indicate that the blinking rates vary for different QDs and for different probe powers, and the quantum dot spends $\sim 80\%$ of the time in the {\it bright} state.

Photon blockade and photon-induced tunneling are quantified by the normalized second-order correlation function $g^{(2)}(\tau,\Delta \omega_{p} / g)$. Each peak in the histogram of Fig.\ref{fig:antibunch} represents the unnormalized value of  $g^{(2)}(\tau,\Delta \omega_{p} / g)$ averaged over the pulse duration of 40 ps. We express this time averaging by using the notation $\overline{g}^{(2)}(\tau,\Delta \omega_{p} / g)$. Because of the QD blinking there are two choices for the normalization constant. One choice is to normalize such that $\overline{g}^{(2)}(\tau \rightarrow \infty, \Delta \omega_{p} / g)=1$. We keep the notation $\overline{g}^{(2)}(\tau,\Delta \omega_{p} / g)$ for this normalized quantity (Fig.\ref{fig:antibunch}(b)). We stress that $\overline{g}^{(2)}(\tau,\Delta \omega_{p} / g)$ captures both the quantum and classical nature (i.e. blinking) of the output field. The normalization constant $\overline{G}^{(2)}_{\infty}$ for $\overline{g}^{(2)}(\tau,\Delta \omega_{p} / g)$ is determined by fitting the histogram with the function $\overline{G}^{(2)}(m T_{0})=(\overline{G}^{(2)}_{0}-\overline{G}^{(2)}_{\infty}) \exp{[-m T_{0}/T]}+ \overline{G}^{(2)}_{\infty}$ for $m \geq 1$. The quantity $\overline{G}^{(2)}(\tau)$ represents the number of counts at time $m T_{0}$, where $m$ indexes the peak number with $m=0$ corresponding to $\tau=0$,  and $T_{0}=12.5$ ns is the pulse repetition period. The other choice of normalization constant is $\overline{G}^{(2)}_{0}$, which is equivalent to normalizing to the nearest neighbor peaks at $\tau = \pm 12.5$ ns. When referring to this normalization we use the notation $\overline{g}_{0}^{(2)}(\tau,\Delta \omega_{p} / g)$ (Fig.\ref{fig:antibunch}(b)). The time-dependence in $g^{(2)}(\tau,\Delta \omega_{p} / g)$ on the time scale of 12.5 ns is due almost exclusively to the fast interaction between probe and QD and is only negligibly affected by the comparatively slow QD blinking process. Thus the nearest-neighbor normalized $\overline{g}_{0}^{(2)}(\tau,\Delta \omega_{p} / g)$ better captures the photon blockade ($\overline{g}_{0}^{(2)}(0,\Delta \omega_{p} / g) < 1$) and photon-induced tunneling ($\overline{g}_{0}^{(2)}(0,\Delta \omega_{p} / g) > 1$) effects.

\begin{figure}[htbp]
    \includegraphics[width=3.5in]{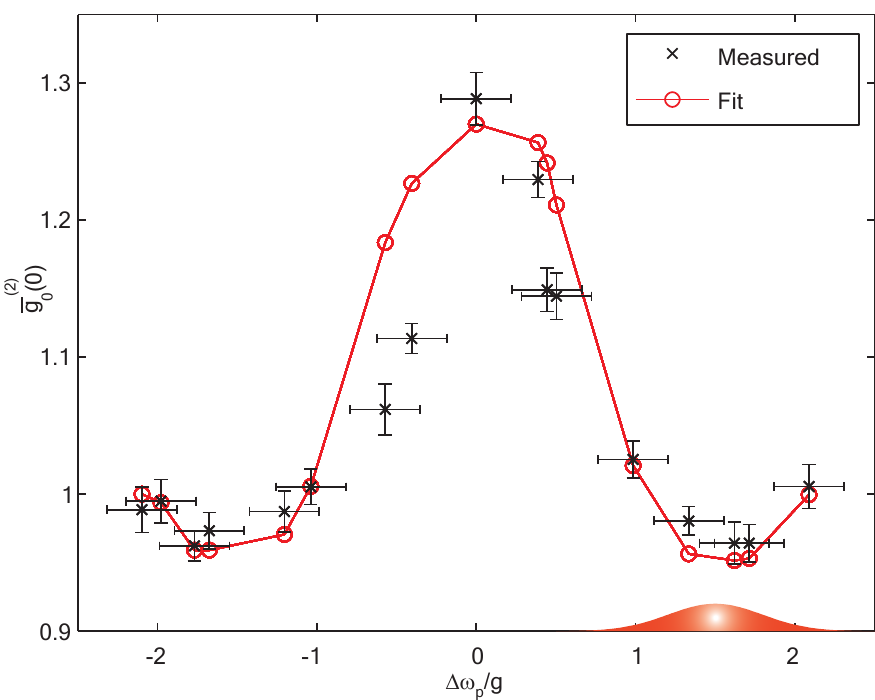}
    \caption{\footnotesize Second-order correlation function $\overline{g}_{0}^{(2)}(0,\Delta \omega_{p} /g)$ for different detunings between the probe and the anticrossing frequency. As the probe is tuned through the resonance of the QD-cavity system, the output field shows antibunched and bunched behavior as it transitions from photon blockade regime to photon-induced tunneling regime. The fit takes into account the finite pulse-width of the probe, QD blinking and background due to the imperfect extinction of the cross-polarized setup. In the bottom-right corner we show the relative width of the pulsed laser probe.}
    \label{fig:g2spec}
\end{figure}

In the case of photon blockade, $\overline{g}^{(2)}(0,1.5) =0.912 \pm 0.005$ and $\overline{g}_{0}^{(2)}(0,1.5)=0.881\pm0.009$ (Fig.\ref{fig:antibunch}(b)), showing the antibunched quantum nature of the system for both choices of normalization. For photon-induced tunneling, $\overline{g}^{(2)}(0,0) =1.33 \pm 0.02$ and $\overline{g}_{0}^{(2)}(0,0)=1.27 \pm 0.02$ (Fig.\ref{fig:antibunch}(d)). These values are different from the ideal theoretical prediction in Fig.\ref{fig:theory}(c) because of the background noise caused by the imperfect extinction of the cross-polarized experimental setup (signal to noise ratio $\sim 6:1$), QD blinking and finite bandwidth of the probe. Both the background and the output signal when the QD is in the {\it dark} state have $\overline{g}^{(2)}(\tau, \Delta \omega_{p} / g)=1$. 

We repeated the autocorrelation measurements for a large set of detunings to map the full spectrum of $\overline{g}_{0}^{(2)}(0,\Delta \omega_{p} /g)$. The measurement of the full autocorrelation spectrum entails several challenges such as sample drift resulting in fluctuating coupling intensity into the cavity, and fluctuating temperature. To map the dependence of $\overline{g}_{0}^{(2)}(0,\Delta \omega_{p} /g)$ on probe detuning, we took care to maintain constant coupling into the cavity mode for the full duration of the experiment. Our most consistent data set is presented in Fig.\ref{fig:g2spec} where we plot $\overline{g}_{0}^{(2)}(0,\Delta \omega_{p} /g)$ for different detunings of the probe frequency. We kept a constant probe power of $\sim 1.0$ nW before the objective lens corresponding to an average cavity photon number $\left\langle n \right\rangle \sim 0.4$, and the coupling was re-optimized for every data point. The lowest value for $\overline{g}_{0}^{(2)}(0,\Delta \omega_{p} /g)$ obtained in this data set is not as antibunched as the value $\overline{g}_{0}^{(2)}(0,1.5)=0.88$ reported in Fig.\ref{fig:antibunch}(b), mainly because we could not reproduce exactly the same coupling conditions. We found that the data in Fig.\ref{fig:g2spec} is well fitted by a numerical model that takes into account pulses of finite bandwidth, QD blinking and background from the imperfect extinction of the cross-polarized setup (see Methods section for details).

The experimental data in Fig.\ref{fig:g2spec} shows that this strongly coupled system allows control of the statistics of the output field from sub-poissonian to super-poissonian. Thus, the probability of transmitting a specific number of photons can be tailored by controlling $\Delta \omega_{p}$.  Given the above-mentioned properties, this device can be used as a single photon transistor\cite{LukinSPhTr} where the presence of a gate field could control the transmission of a signal field. One scheme to implement the transistor is to tune the frequency of the gate field resonant with one of the polaritons in the first-order manifold, say $\omega_{0}+g$. A photon injected at $\omega_{0}+g$, increases the probability of absorbing photons that are resonant with the $\ket{1,+} \rightarrow \ket{2,+}$ transition at $\omega_{0}+g(\sqrt{2}-1)$. If the signal is tuned to this frequency, the presence of the gate field enhances the transmission of the signal field. The photonic crystal architecture allows for easy integration of such a single photon transistor with photonic crystal waveguides\cite{AndreiWgCoupling,Noda2DTrap} so the single photon switching is done directly on the chip. The most straightforward configuration would be a photonic crystal cavity butt-coupled in between two photonic crystal waveguides\cite{NotomiTrapDelay}.

In conclusion we observed the phenomena of photon blockade and photon-induced tunneling in a solid-state strongly coupled QD/cavity system. By changing the frequency of the input beam, we synthesized nonclassical states of light with various degrees of bunching or antibunching ($0.91 < \overline{g}^{(2)}(0) < 1.34$). These results complement the photon blockade experiments in atomic physics by showing the full dependence of $g^{(2)}(\tau)$ on the probe frequency. They have  promise in realizing single photon transistors for on-chip optical logic devices operating at the single photon level, or the implementation of two-qubit gates and generation of nonclassical states for quantum information processing, quantum lithography and quantum metrology.

\section*{Methods}

\begin{small}

{\bf Autocorrelation Measurement:} We scan several cavities until we find one which contains a strongly coupled QD, as determined by the anticrossing behavior in photoluminescence between QD and cavity during temperature tuning.  Then we direct the pulsed laser beam at the cavity and observe the reflected beam in cross-polarization.  While tuning the local temperature with an additional heating beam, we adjust the probe beam coupling to optimize the QD-induced reflectivity drop, as described for the continuous-wave beam in ref.\cite{NatureRef}.  Then we stop scanning and temperature-tune the QD and cavity onto resonance.  With the pulsed probe beam at different detunings with respect to the anticrossing point, we measure the autocorrelation signal by passing the reflected probe through a grating filter (to remove stray light) followed by the HBT setup. To limit sample drift, the alignment procedure is repeated for every data point in Fig. 4.\linebreak

{\bf Data Analysis:} The numerical model for the second-order coherence in Fig.4 is based on numerical integration of the quantum master equation.  A time-dependent driving term in the Hamiltonian represents the 40-ps excitation pulses.  The intensity of the drive field matches the intensity used in the experiment, representing one-third of saturation. In our experiment, this intensity was $1$ nW for the incident beam, measured before the objective lens. The state of the QD/cavity is time-evolved using a quantum Monte Carlo approach, which we based on the qotoolbox of ref.\cite{TanMATLAB}.  The QD can emit into free space or into the cavity mode, which in turn dissipates energy into the output channel at the loss rate $\omega/Q$.  We then compute the autocorrelation on the output channel. The simulation additionally accounts for QD blinking and laser background. The full second-order coherence is calculated as weighted sum of the different contributions, 
\begin{eqnarray}
G^{(2)}(\tau)&=&\braket{\oad(t)\oad(t+\tau) \oa(t+\tau)\oa(t) \rho}\\
&=& p_{B} G^{(2)}_{B}(\tau)+p_{BG} G^{(2)}_{BG}(\tau)+p_{D} G^{(2)}_{D}(\tau)
\end{eqnarray}
where the autocorrelation function $G^{(2)}_B(\tau)$ accounts for the QD bright state, $G^{(2)}_D(\tau)$ for the QD-dark state (calculated using $g\rightarrow 0$), and $G^{(2)}_{BG}(\tau)$ for background laser signal (a coherent state). We then normalize by the simulated nearest neighbor peaks ($\overline{G}^{(2)}_{0}$) to obtain $g_{0}^{(2)}(\tau)$, as shown for example in Fig.\ref{fig:antibunch}(d).

\end{small}

\section*{Acknowledgements}

Financial support was provided by the MURI Center for photonic quantum information systems (ARO/IARPA Program supervised by Dr. TR Govindan)), ONR Young Investigator Award( supervised by Dr. Chagaan Baatar) and NSF. D.E and I.F. were also supported by the NDSEG fellowship. Part of the work was performed at the Stanford Nanofabrication Facility of NNIN supported by the National Science Foundation.


\bibliographystyle{unsrt}

\end{document}